\documentclass[aps,reprint,showpacs,floatfix,citeautoscript,longbibliography,superscriptaddress]{revtex4-2}
\usepackage{graphicx}
\usepackage{times}
\usepackage{bm,amsmath,amssymb,mathrsfs,dcolumn}
\usepackage[colorlinks=true, linkcolor=blue, citecolor=blue, urlcolor=blue, linktoc=page, bookmarks=false]{hyperref} 
\usepackage[figure]{hypcap}
\usepackage[usenames]{color}
\hypersetup{pdfborder=1 1 1,colorlinks=true,citecolor=blue,linkcolor=blue}
\usepackage{epstopdf}
\usepackage{subfigure}
\usepackage{units}
\usepackage{breakurl}
\usepackage{multirow}
\usepackage{graphicx}
\usepackage{booktabs}

\usepackage{ulem}

\begin{document}

\title{Crystal facet orientated Altermagnets for detecting ferromagnetic and antiferromagnetic states by giant tunneling magnetoresistance effect}
\author{Boyuan Chi}
\affiliation{Beijing National Laboratory for Condensed Matter Physics, Institute of Physics, University of Chinese Academy of Sciences, Chinese Academy of Sciences, Beijing 100190, China}
\affiliation{Center of Materials Science and Optoelectronics Engineering, University of Chinese Academy of Sciences, Beijing 100049, China}

\author{Leina Jiang}
\affiliation{Beijing National Laboratory for Condensed Matter Physics, Institute of Physics, University of Chinese Academy of Sciences, Chinese Academy of Sciences, Beijing 100190, China}

\author{Yu Zhu}
\affiliation{Beijing National Laboratory for Condensed Matter Physics, Institute of Physics, University of Chinese Academy of Sciences, Chinese Academy of Sciences, Beijing 100190, China}

\author{Guoqiang Yu}
\affiliation{Beijing National Laboratory for Condensed Matter Physics, Institute of Physics, University of Chinese Academy of Sciences, Chinese Academy of Sciences, Beijing 100190, China}
\affiliation{Songshan Lake Materials Laboratory, Dongguan, Guangdong 523808, China}

\author{Caihua Wan}
\affiliation{Beijing National Laboratory for Condensed Matter Physics, Institute of Physics, University of Chinese Academy of Sciences, Chinese Academy of Sciences, Beijing 100190, China}

\author{Jia Zhang}
\altaffiliation{jiazhang@hust.edu.cn}
\affiliation{School of Physics and Wuhan National High Magnetic Field Center, Huazhong University of Science and Technology, 430074 Wuhan, China}

\author{Xiufeng Han}
\altaffiliation{xfhan@iphy.ac.cn}
\affiliation{Beijing National Laboratory for Condensed Matter Physics, Institute of Physics, University of Chinese Academy of Sciences, Chinese Academy of Sciences, Beijing 100190, China}
\affiliation{Center of Materials Science and Optoelectronics Engineering, University of Chinese Academy of Sciences, Beijing 100049, China}
\affiliation{Songshan Lake Materials Laboratory, Dongguan, Guangdong 523808, China}


\begin{abstract}
  Emerging altermagnetic materials with vanishing net magnetizations and unique band structures have been envisioned as an ideal electrode to design antiferromagnetic tunnel junctions. Their momentum-resolved spin splitting in band structures defines a spin-polarized Fermi surface, which allows altermagnetic materials to polarize current as a ferromagnet, when the current flows along specific directions relevant to their altermagnetism. Here, we design an Altermagnet/Insulator barrier/Ferromagnet junction, renamed as altermagnetic tunnel junction (ATMTJ), using RuO$_2$/TiO$_2$/CrO$_2$ as a prototype. Through first-principles calculations, we investigate the tunneling properties of the ATMTJ along the [001] and [110] directions, which shows that the tunneling magnetoresistance (TMR) is almost zero when the current flows along the [001] direction, while it can reach as high as 6100\% with current flows along the [110] direction. The spin-resolved conduction channels of the altermagnetic RuO$_2$ electrode are found responsible for this momentum-dependent (or transport-direction-dependent) TMR effect. Furthermore, this ATMTJ can also be used to readout the N\'{e}el vector of the altermagnetic electrode RuO$_2$. Our work promotes the understanding toward the altermagnetic materials and provides an alternative way to design magnetic tunnel junctions with ultrahigh TMR ratios and robustness of the altermagnetic electrode against external disturbance, which broadens the application avenue for antiferromagnetic spintronic devices.

\end{abstract}

\maketitle

  Recently, the altermagnetism, as a unique type of antiferromagnetism with momentum-locked spin splitting and alternating spin polarizations in both real-space (crystalline structure) and momentum-space (band structure) \cite{PRX-12-031042-2022,PRX-12-040501-2022} have attracted immense attention owing to some desired performances \cite{PRX-12-031042-2022,PRX-12-040501-2022,PCCP-2016,JPSJ-2019,PRB-102-2020,PRM-2021,SA-2020,PRX-12-011028-2022}. An altermagnetic material has zero net magnetization, just like an antiferromagnetic (AFM) material; however, non-relativistic spin splitting occurs on some high-symmetric lines in the band structure of the altermagnetic material, which allows this special AFM material to behave similarly as a ferromagnetic (FM) counterpart in some cases. A certain magnetic space groups (MSGs) with the violated $TP\tau$ and $U\tau$ symmetries support this non-relativistic spin splitting, where $T$, $P$, $U$, $\tau$ stand for the time reversal, spatial inversion, spinor symmetry and the half lattice translation\cite{JPSJ-2019,PRM-2021}. Since magnetism has already been enriched by the fascinating altermagnetism, it is thus a natural generalization to upgrade the most important spintronic device, magnetic tunnel junctions (MTJs) \cite{S-2001,RMP-2004,IEEE-2006}, with altermagnetic electrodes to realize a higher tunneling magnetoresistance (TMR) effect (if possible) and verify the theoretically promised features of altermagnetic materials.
  
  MTJs consisting of the FM/Insulator barrier/FM sandwich structure \cite{PLA-1975,PRL-74-1995,JMMM-1995} act as the elementary core of the spintronic devices in such practical applications as nonvolatile magnetic random-access memories (MRAMs) and magnetic sensors \cite{JPCM-15-R109-2003,JPCM-15-R1603-2003}. By changing the magnetization alignment of two FM electrodes, antiparallel or parallel, an MTJ can be switched between its high and low resistance states, which is well known as the TMR effect and functions as the working principle of the MTJ devices. On the other hand, AFM materials have long been realized advantageous in the following aspects than FM materials, no stray magnetic field and robustness against external fields, strong anisotropy, ultrafast spin dynamics, attractive for antiferromagnetic spintronics. However, according to the classic viewpoint, lack of global spin polarization hampers an AFM from being adopted as an electrode in MTJ. Lately, some pioneers have theoretically and experimentally verified that an all-antiferromagnetic tunnel junction (AFMTJ) can also generate the TMR effect, by using two altermagnetic electrodes such as RuO$_2$ \cite{NC-2021,PRX-12-011028-2022}, Mn$_3$Sn \cite{PRL-128-2022,N-613-490-2023} and Mn$_3$Pt \cite{N-613-485-2023} owing to an emerging “local” spin polarization in their momentum space.
  
  Compared to sophisticated MTJs, an AFMTJ made of Mn$_3$Pt can also exhibit as high TMR ratio as ~100\% at room-temperature \cite{N-613-485-2023} and maintain the merit of vanishing stray field. However, currently, magnetic and electrical methods to control the 180$^{\circ}$ reversal of the Néel vector of AFM materials are still relatively limited. Taking RuO$_2$/TiO$_2$/RuO$_2$ collinear AFMTJ as an example, it is difficult for a magnetic field to switch the N\'{e}el vector of the RuO$_2$ free layer. This deficiency in the controlling method thus makes (1) an unambiguous determination of the altermagnetism in some AFM material challenging and (2) an all AFM-based AFMTJ incapable of developing magnetic sensors in despite of a predicted high TMR ratio.
  
  Instead, by replacing one AFM electrode of an AFMTJ with a FM electrode, we propose an Altermagnet/Insulator barrier/FM tunnel junction (renamed as altermagnetic tunnel junction or ATMTJ for short) via utilizing the unique band structure of the altermagnet. From the analysis of the Fermi surface, it is found that the altermagnet can polarize a current when transportation occurs along some specific crystal directions, just like a FM electrode though with zero net magnetization. The TMR effect is generated by switching the magnetization of the FM electrode, which however can be straightforwardly used as a strong proof to testify the “local” spin polarization of the altermagnetic AFM. Moreover, in traditional MTJ spin valves, an additional pinned layer by an AFM layer is required to fix the magnetization of the reference layer, and the magnetizations of the pinned and reference layers need antiparallelly aligned and, more favorably, fully compensated to reduce the influence of their stray fields on the free layer. This hybrid structure containing the pinned layer, the reference layer and a nonmagnetic spacer, is famed as the synthetic antiferromagnet (SAF). However, since an altermagnetic electrode has zero net magnetization and its N\'{e}el order is robust against an external magnetic field, a single layer of the altermagnet can conveniently act both as a pinned layer and a reference layer, the natural absence of whose stray field is benefit to improve linearity and reduce noise of as-based magnetic sensors. From this perspective, a reliable ATMTJ candidate with matched lattice parameters, epitaxial growth capability and high TMR is urgently needed to be given from theoretical predictions.
  
  Using RuO$_2$/TiO$_2$/CrO$_2$ ATMTJ as a prototype, we predict the transport properties of the ATMTJ along the [001] and [110] crystal orientations through the first-principles calculations as described in the Supplemental Material \cite{SM}. The altermagnetic RuO$_2$ has the Fermi surface with momentum-dependent non-relativistic spin splitting. Thus, if a current flows along the [001] direction of RuO$_2$, the shape of the two spins conduction channels is the same, resulting in a spin-neutral current, which cannot produce the TMR effect with the FM electrode CrO$_2$. But if a current flows along the [110] direction of RuO$_2$, the conduction channels of two spins are completely different, resulting in a spin-polarized current which generates a sizeable TMR effect of 6100\% with the FM CrO$_2$ electrode. Thus, this ATMTJ is an ideal candidate for magnetic sensors because of the AFM nature of the RuO$_2$ reference layer, which can avoid disturbance of a magnetic field on the reference layer. Furthermore, this RuO$_2$/TiO$_2$/CrO$_2$ ATMTJ also provides a method to readout the N\'{e}el vector of RuO$_2$ with the (110) crystalline plane, facilitating the wider applications of the altermagnetic AFM materials.

\begin{figure}
  	\centering
  	\includegraphics[width=1.0\linewidth]{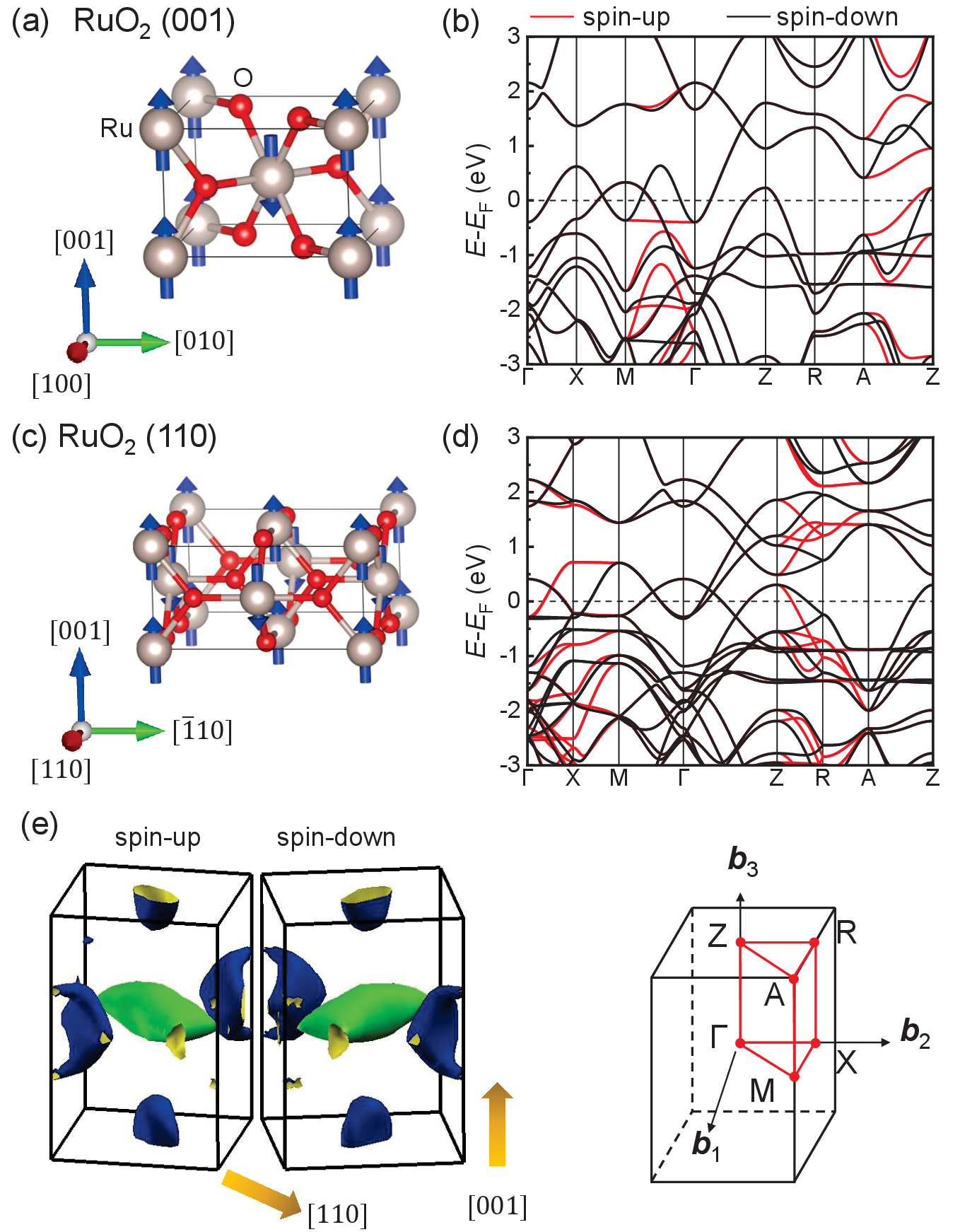}
  	\caption{The atomic and magnetic structures of RuO$_2$ with (a) (001) and (c) (110) crystal face. The band structures of RuO$_2$ with (b) (001) and (d) (110) crystal face. (e) The spin-up and spin-down Fermi surfaces of RuO$_2$ (plotted using the xcrysden package\cite{xcry}). The orange arrows indicate the transport direction in the proposed ATMTJs.}
  	\label{fig1}
  \end{figure}

  Bulk rutile RuO$_2$ is a collinear AFM with $P4_2/mnm$ space group and $P4_{2}{'}/mnm'$ MSG as depicted in Fig. \hyperref[fig1]{1(a)}. The lattice parameters of the relaxed RuO$_2$ are $a$=$b$=4.49 \AA{}, $c$=3.12 \AA{}, and the magnetic moment of Ru atom is 1.2 $\mu_{\rm B}$, which is consistent with the previous works \cite{AC-1997,PRL-118-2017}. Since its MSG violates the $TP\tau$ and $U\tau$ symmetries, RuO$_2$ has a nontrivial spin splitting band structure that depends on momentum $\boldsymbol{k}$ \cite{SA-2020,PRL-128-2022}. As shown in Fig. \hyperref[fig1]{1(b)}, the spin-up and spin-down bands of RuO$_2$ split along the $\Gamma$-M and Z-A directions, while they degenerate at the other high-symmetry lines. This $\boldsymbol{k}$-dependent spin polarization is envisioned promisingly to design spintronic devices along those spin-nontrivial directions.
  
  The Fermi surface of an electrode determines the distribution of conductive electrons with different spins in the momentum space, which plays a decisive role to the TMR ratio of the MTJs. As depicted in Fig. \hyperref[fig1]{1(e)}, the spin-up and spin-down Fermi surfaces of RuO$_2$ share the same shape, and can be easily converted to each other by some rotational operations. This Fermi surface indicates that if electrons transport along the [001] direction of RuO$_2$, a zero spin-polarized current will be generated. However, if electrons transport along $\Gamma$-M ($i.e.$, [110] direction), the current will be polarized, because the spin-up and spin-down Fermi surfaces of RuO$_2$ have diverse projections on the (110) plane. Therefore, RuO$_2$, though as an antiferromagnet, can still act as a spin polarizer, like a FM electrode, when a current flowing along this [110] direction. Fig. \hyperref[fig1]{1(c)} is the atomic structure of RuO$_2$ cleaved from the (110) crystalline plane. The two spin subbands of RuO$_2$ (110) split along the $\Gamma$-X, X-M, Z-R and R-A directions, as shown in Fig. \hyperref[fig1]{1(d)}.
  
  \begin{figure}
	\centering
	\includegraphics[width=1.0\linewidth]{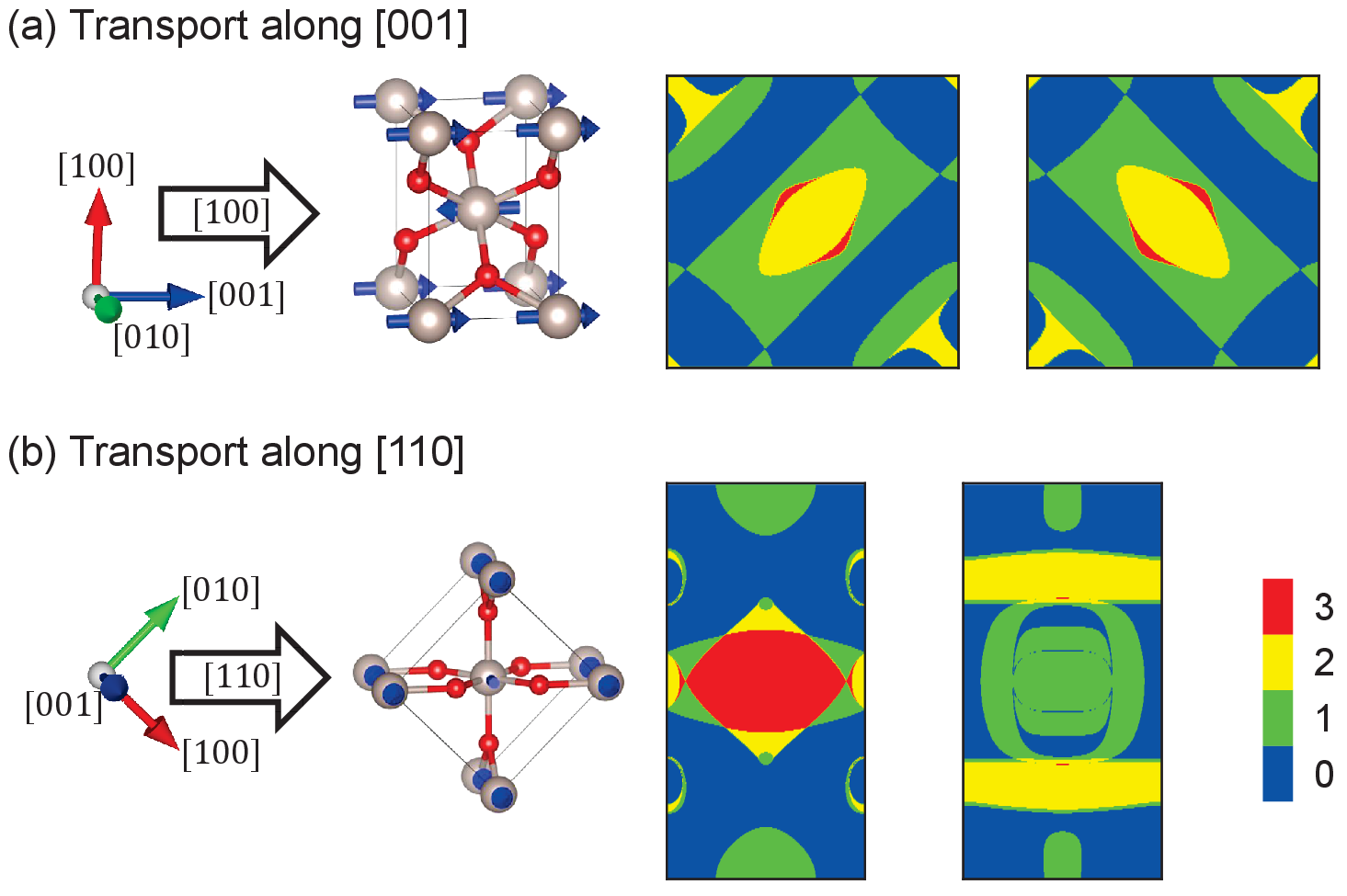}
	\caption{The spin- and $\boldsymbol{k}_{\parallel}$-resolved conduction channels of the bulk RuO$_2$ transport along the (a) [001] and (b) [110] directions. Upper and lower panels are the spin-up and spin-down channels, respectively.}
	\label{fig2}
  \end{figure}

\begin{figure}[b]
  	\centering
  	\includegraphics[width=1.0\linewidth]{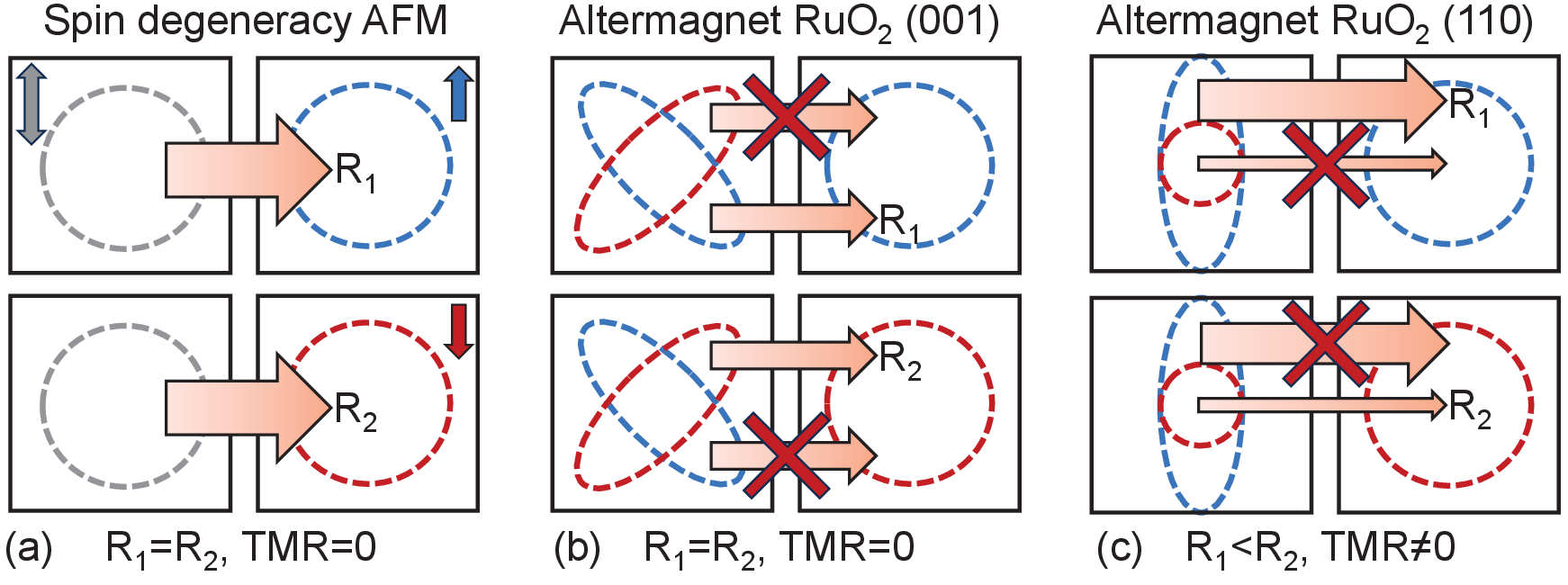}
  	\caption{Schematics of (a) spin-degenerate AFM/barrier/FM, (b) altermagnetic RuO$_2$ (001)/barrier/FM and (c) altermagnetic RuO$_2$ (110) tunnel junctions. Grey indicates spin-up or spin-down conduction channels of conventional AFMs. Blue and red indicate the spin-up and spin-down conduction channels of altermagnetic AFMs, respectively. For simplicity, magnetically half-metallic material is selected as the FM electrode.}
  	\label{fig3}
  \end{figure}

  To provide a clearer explanation on the properties of RuO$_2$ along different transport directions, the conduction channels of RuO$_2$ (001) and RuO$_2$ (110) are plot in Fig. \hyperref[fig2]{2}. The conduction channels are the available number of propagating Bloch states in the momentum space, which determines the ballistic transmission. It can be straightforwardly regarded as the projection of the Fermi surface on the plane perpendicular to the transport direction. When a current flows along the [001] direction of RuO$_2$ (001), the conductance channels of the spin-up and spin-down electrons have the same distribution with only a rotational difference by a 90$^{\circ}$ angle [Fig. \hyperref[fig2]{2(a)}], causing the current passing through RuO$_2$ (001) not polarized. Thus, if RuO$_2$ and a traditional FM are used as two electrodes of an ATMTJ and a current flows along the [001] direction, it is impossible to generate the TMR effect, because the transmission under different magnetization arrangements is the same, as schematically shown in Fig. \hyperref[fig3]{3(b)}. On the other hand, when a current flows along the [110] direction of RuO$_2$ (001), the conduction channels of two spins are completely different, as Fig. \hyperref[fig2]{2(b)} revealed. Note that since the in-plane lattice constants of RuO$_2$ (110) are $a$ = $4.49\times\sqrt{2}$ \AA{}, $c$ = 3.12 \AA{}, its $\boldsymbol{k}$-space has a rectangular shape. Hence, the TMR effect can in principle emerge when the RuO$_2$ and a traditional FM is used to compose an ATMTJ with the (110) plane [Fig. \hyperref[fig3]{3(c)}].
  
  Then, we used RuO$_2$/TiO$_2$/CrO$_2$ ATMTJ as a prototype to analyze the transport properties of Altermagnet/Insulator/FM tunnel junctions. Notably, the CrO$_2$ electrode is chosen here just for lattice matching as shown below, which facilitates the $ab$-$initio$ calculation. CrO$_2$ is not exclusive, in fact, other ferromagnetic electrodes with highly spin-polarized Fermi surface such as Fe, CoFe, CoFeB $etc.$, should also be applicable. Therefore, in this ATMTJ, only RuO$_2$ is required to be a single crystal, which is easier for industrial production than AFMTJs whose two AFM electrodes should both have the required single-crystalline structures. Here, TiO$_2$ and CrO$_2$ both have the rutile structure, the relaxed lattice parameters of TiO$_2$ (CrO$_2$) are $a$=$b$=4.60 \AA{}, $c$=2.95 \AA{} ($a$=$b$=4.41 \AA{}, $c$=2.90 \AA{}), which matches well with the experimental values \cite{AM-1995,PRB-88-2013}. The atomic structure of CrO$_2$ is shown in Supplemental Material \cite{SM} Fig. S1(a,d), its band structure indicates its magnetically half-metallic characteristic [Fig. S1(b,e)]. Such a FM electrode with a high spin polarization can significantly increase the TMR effect of the composed ATMTJs. Figs. S1(c,f,g) depict the conduction channels and the Fermi surface of CrO$_2$.
  
  We first constructed the RuO$_2$ (001)/TiO$_2$/CrO$_2$ ATMTJ with electrons transporting along the [001] direction, as shown in Supplemental Material \cite{SM} Fig. S2(a). RuO$_2$ is the reference layer with its N\'{e}el vector fixed along the [001] direction and CrO$_2$ is the ferromagnetically free layer. When the magnetization of CrO$_2$ is parallel to the N\'{e}el vector of RuO$_2$, it is defined as the P state of the ATMTJ, and when the magnetization is anti-parallel to the N\'{e}el vector, it is defined as the AP state. From the above symmetric analysis, it can be inferred that this ATMTJ can hardly produce the TMR effect, because the matching (or overlap) of the CrO$_2$ with the two spin-resolved conduction channels of RuO$_2$ is equivalent. Fig. S2(c) shows that the transmission coefficient distribution of the P state is the same with that of the AP state. As a result, these two states have very close conductance, leading to an almost zero TMR ratio [see Supplementary Material \cite{SM} Table S I.]. Note that the non-zero TMR ratio can be attributed to the slightly uncompensated magnetic moment of RuO$_2$ at the interface [Fig. S2(b)], where the spin-down moment is slightly bigger than that of the spin-up (about 0.1 $\mu _{\rm B}$), resulting in a little higher conductance of the AP state with spin-down tunneling. Therefore, this ATMTJ with the [001] transport direction cannot exhibit a sizable TMR effect.
  
\begin{figure}
	\centering
	\includegraphics[width=1.0\linewidth]{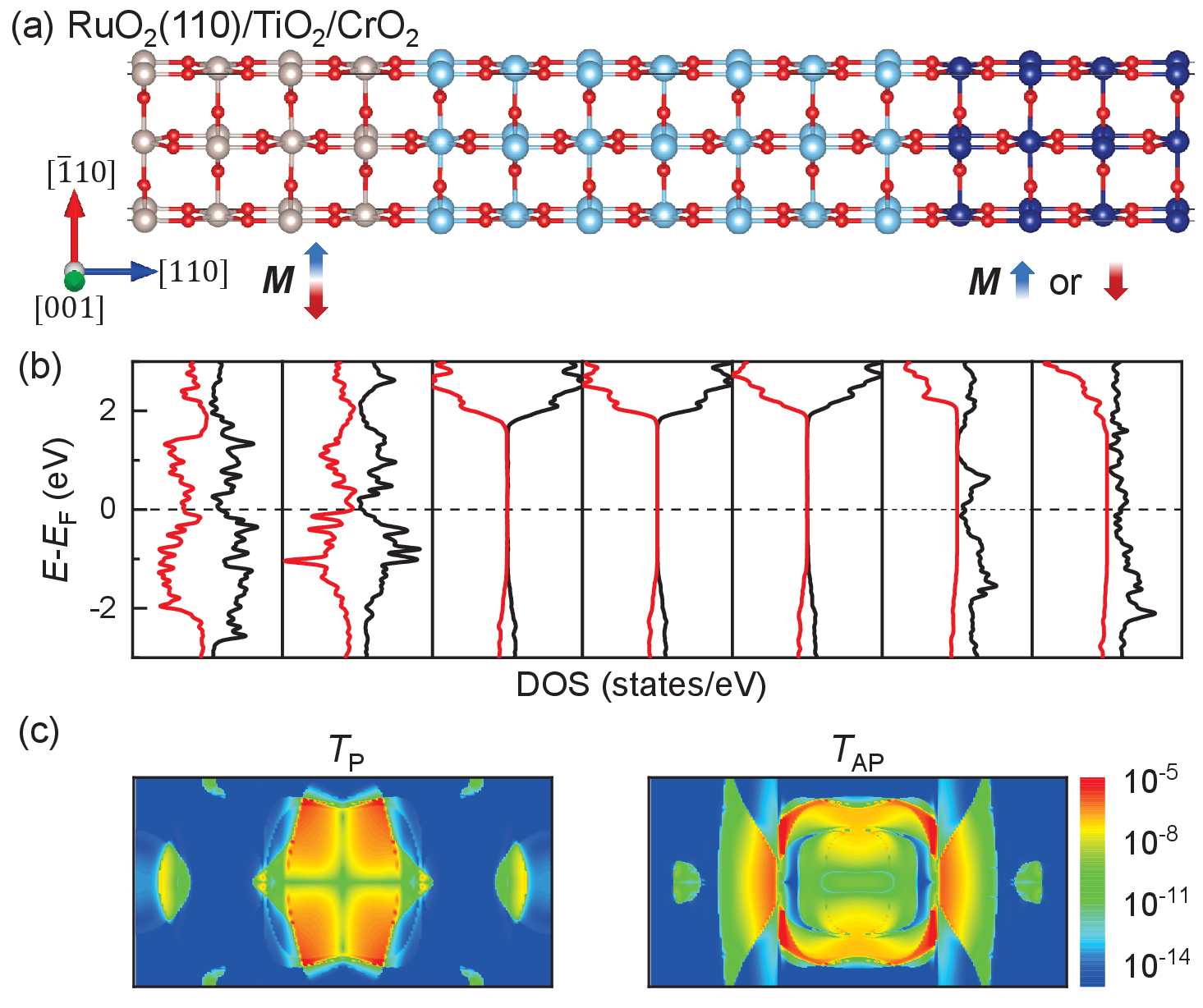}
	\caption{(a) The atomic structure of RuO$_2$ (110)/TiO$_2$/CrO$_2$ ATMTJ. (b) The spin- and layer-resolevd density of states for this ATMTJ. (c) The $\boldsymbol{k}_{\parallel}$-resolved transmission coefficient distributions in RuO$_2$ (110)/TiO$_2$/CrO$_2$ ATMTJ with P and AP magnetization configurations.}
	\label{fig4}
\end{figure}
  
   Then we constructed the RuO$_2$ (110)/TiO$_2$/CrO$_2$ ATMTJ with electrons transporting along the [110] direction, as shown in Fig. \hyperref[fig4]{4(a)}. The definition of the two magnetization configurations (P and AP) of the ATMTJ is consistent with above. The calculated conductance difference between the two magnetization configuration states is nearly two orders of magnitude, resulting in a large TMR ratio of 6100\% [see Table S I.]. Fig. \hyperref[fig4]{4(b)} is the layer-resolved density of states of the ATMTJ, the net magnetization of the interface RuO$_2$ is very small ($<$ 0.1 $\mu_{\rm B}$), which cannot generate such huge TMR effect. As depicted in Fig.  \hyperref[fig4]{4(c)}, it can be clearly identified that the transmissions of the two states are quite different, and the distributions of transmission coefficients are similar to the overlapping part of the conduction channels of RuO$_2$ (110) and CrO$_2$ (110). Consequently, the RuO$_2$ (110)/TiO$_2$/CrO$_2$ ATMTJ can produce a large TMR effect by switching the magnetization of the CrO$_2$ electrode, and meanwhile avoid any disturbance from an external magnetic field on the Néel vector of the RuO$_2$ reference layer. This characteristic makes the ATMTJ ideally prospective for the field of magnetic sensors. Furthermore, this ATMTJ can be suitably used to readout the N\'{e}el vector of the RuO$_2$ (110). The field-dependence of the resistance of the ATMTJ will be completely reversed when the Néel vector of RuO$_2$ (110) is switched from 0$^{\circ}$ to 180$^{\circ}$ or vice versa, as schematically shown in Supplemental Material \cite{SM} Fig. S3.

  The transport properties of the RuO$_2$ (110)/TiO$_2$/CrO$_2$ ATMTJ discussed above were calculated at the Fermi energy. To further investigate the influence of doping, strain, deficiency, $etc.$ induced Fermi level shift (if any) on the TMR ratio, we calculated the transmission and TMR ratio at different energies near the Fermi level, as shown in Supplemental Material \cite{SM} Fig. S4. It is found that there is a sign inversion of the TMR ratio when the energy approaches $E_{\rm F}$+0.04 eV, which can be explained by the conduction channels of RuO$_2$ (110) as a function of energy [Fig. S5]. Fig. S4(a) also indicates that the TMR ratio does not vary significantly with energy and remains on the order of about 10$^3$\%.

  Worth noting, if the method to electrically manipulate the magnetization of the AFM electrode in the AFMTJ is developed and the epitaxial growth technology to fabricate all single-crystalline tunnel junctions is mature in the future, another RuO$_2$ (110)/TiO$_2$/RuO$_2$ AFMTJ can be constructed along the (110) crystalline plane. We calculated this AFMTJ by replacing the CrO$_2$ in Fig. \hyperref[fig4]{4} with RuO$_2$, and obtained a TMR ratio of 685.1\%. The AFMTJs possess the advantages of no stray field and ultrafast spin dynamics \cite{NN-2016,RMP-2018}, which are also valued in the field of spintronic devices.

  It is important to mention that RuO$_2$ is not indispensable to construct the ATMTJ, any antiferromagnetic metal with spin-polarized Fermi surfaces can serve as an alternative. CrSb is another altermagnet with N\'{e}el temperature high up to 705 K \cite{PR-2008}, whose magnetic space group is $P6_{3}{'}/m'm'c$. As shown in Supplementary Material \cite{SM} Fig. S6, CrSb exhibits a spin splitting along $\Gamma$-L and $\Gamma$-L'. From the analysis of Fermi surface and conduction channel in Figs. S7 and S8, it can be seen that the current will be polarized as it flows along the [10$\overline{1}$1] direction of the CrSb, but remains spin-neutral when it flows along other directions. Therefore, CrSb can also served as antiferromagnetic electrode in the ATMTJ, exhibiting similarly remarkable TMR effect. With the development of the altermagnetism research in the future, more and more materials for constructing altermagnet electrode in ATMTJ will be promisingly discovered.

   In conclusion, through first-principles calculations, we proposed a kind of Altermagnet/Insulator/FM ATMTJ whose TMR ratio is dependent on the transport direction. The energy band of the altermagnetic material appears spin splitting along certain directions, so when a current flows along the high symmetry direction, the conduction channels of spin-up and spin-down have the same shape, which cannot generate TMR effect with FM material. However, when a current flows along the low symmetry direction, the two spins conduction channels are completely different, which can produce TMR effect with FM material. By using RuO$_2$/TiO$_2$/CrO$_2$ ATMTJ as an example, we obtained almost zero TMR ratio when transporting along the [001] direction and a large TMR ratio when transporting along the [110] direction. This transport properties are related to the nontrivial band structure of RuO$_2$. Since its MSG breaks the $TP\tau$ and $U\tau$ symmetries, the two spins energy bands of AFM material RuO$_2$ split along some high-symmetry lines. Therefore, the shape of the Fermi surface of the two spins of RuO$_2$ is the same. The conduction channels of the two spins along the [001] direction have an identical shape and can be converted to each other through a 90$^{\circ}$ rotation, while they are completely different along the [110] direction. As a result, RuO$_2$ (110) can be treated as a FM-like electrode, and a huge TMR ratio of 6100\% can be obtained in theory for the RuO$_2$ (110)/TiO$_2$/CrO$_2$ ATMTJ. This ATMTJ can be used as a readout method for the N\'{e}el vector of RuO$_2$. Furthermore, due to the fact that one of the electrodes is ferromagnetic, the difficulty in switching a collinear AFM of a all-AFM MTJ can be avoided. Because of the reference layer is AFM, when a magnetic field rotates the free layer of the ATMTJ, the reference layer is intrinsically robust to the field, making this kind of ATMTJ very suitable for the magnetic sensor applications. In addition, in this kind of ATMTJ, only the altermagnet layer need to be a single crystal, which is easier to grow than AFMTJ, making it convenient to industrial production. Our work also offers a promising way to design ATMTJ with altermagnetic materials and promotes the development and application of altermagnetic spintronics.
   
\begin{acknowledgments}
  This work was financial supported by the National Key Research and Development Program of China [MOST Grant No. 2022YFA1402800], the National Natural Science Foundation of China [NSFC, Grant No. 51831012, 12134017, 11974398 and 12204517]. Jia Zhang was supported by the National Natural Science Foundation of China [NSFC, Grant No. 12174129]. The atomic structures were produced using VESTA software \cite{JAC-2011}. High-performance computing resources for contributing to the research results were provided by Beijing PARATERA Technology Co., LTD.
\end{acknowledgments}

\bibliography{main}

\end{document}